\documentclass[runningheads,psfig]{svmult}
\usepackage{psfig}
\usepackage{makeidx}   
\usepackage{graphicx}  
\usepackage{subeqnar}  
\usepackage{multicol}  
\usepackage{physprbb}  
\makeindex             

\begin{document}
\title*{Radio Observations of High Redshift Star Forming Galaxies}
\toctitle{Radio Observations of High Redshift Star Forming Galaxies}
\titlerunning{Radio Observations of High $z$ Star Forming Galaxies}
\author{C.L. Carilli\inst{1}}
\authorrunning{Carilli}
\institute{NRAO, Socorro, NM, 87801, USA}
\maketitle              
\begin{abstract}
I summarize recent results from
radio observations of high redshift star forming
galaxies, discuss radio continuum emission as a measure of star
formation rate, and consider future capabilities at cm to IR
wavelengths.  
\end{abstract}

\section{Radio surveys to $\mu$Jy sensitivity}

Source counts based on low frequency surveys of the sky with Jy
sensitivity showed a significant departure from 
a Euclidean, non-evolving source population, 
indicating, for the first time, cosmic evolution in a 
source population. The source population entailed luminous 
radio galaxies, with spectral luminosities at 178 MHz:
$\rm P_{178} > 10^{32} ~erg~ s^{-1}~ Hz^{-1}$ \cite{ryle,jauncey}. 
In these sources the synchrotron  radio emission 
is from high energy electrons accelerated in
a relativistic jet emanating from the active galactic  nucleus
(AGN)\cite{bartel}.

Subsequent observations with sub-mJy sensitivity, starting with the
WSRT and continuing with the VLA \cite{windhorst}, revealed 
flattening of the  source counts below 5 mJy (Fig. 1). Windhorst et
al. hypothesized that this flattening was due to a new 
population of sources, namely star forming galaxies with
$\rm P_{178}  < 10^{31} ~erg~ s^{-1}~ Hz^{-1}$.
The radio emission in these sources is from relativistic
electrons accelerated in  supernovae remnant shocks\cite{condon}. 

There has been a recent revival of deep radio surveys, motivated in
large part as follow-up to deep optical, infrared, and (sub)mm
surveys \cite{richards,hopkins,garrett,owen}.  The frequency of
choice is 1.4 GHz for 
these deep surveys, allowing for $\mu$Jy sensitivity with arcsecond
resolution and a wide field of view (FWHM = 30$'$). In the coming
years the Expanded VLA (EVLA) will push to the sub-$\mu$Jy level,
while in 
the coming decades the Square Kilometer Array (SKA) will potentially
probe nJy sources.  As pointed out in numerous papers in these proceedings
(Bertoldi, Adelberger, Hughes, Sanders), radio observations play an
important, complimentary role to observations at other wavelengths, in
that:  
\begin{itemize}
\item They are not plagued by extinction corrections.

\item They provide a rough  estimate of source redshifts, in
combination with (sub)mm observations \cite{yun99,yun00}.

\item Low order CO transitions redshift into the cm bands, revealing
large reservoirs of less dense, cooler gas
\cite{papadopoulos,menten}. This topic will not be discussed herein. 

\item They provide arcsecond astrometry and imaging, thereby 
avoiding the confusion problems inherent in
deep searches for optical counterparts of (sub)mm sources discovered
in low resolution single dish bolometer array surveys.
\end{itemize}

\noindent For example, at the optical limits of the HDF ($I_{AB} < 29$)
one expects, by chance,
three faint galaxies within the 6$''$ error circle of a
typical SCUBA or MAMBO source\cite{lilly}. At 1.4 GHz the 
source counts between 40 $\mu$Jy and 1 mJy obey:
$\rm N(> S_{1.4}) = 2.2\times10^{-6}~ S_{1.4}^{-1.4}$~ arcsec$^{-2}$,
with S$_{1.4}$ in mJy\cite{richards}. The number of  spurious 
$\rm S_{1.4} \ge 22 \mu$Jy sources within the error circle is only
0.02.  

\begin{figure}
\vskip -0.1 in
\hspace*{-0.2in}
\psfig{figure=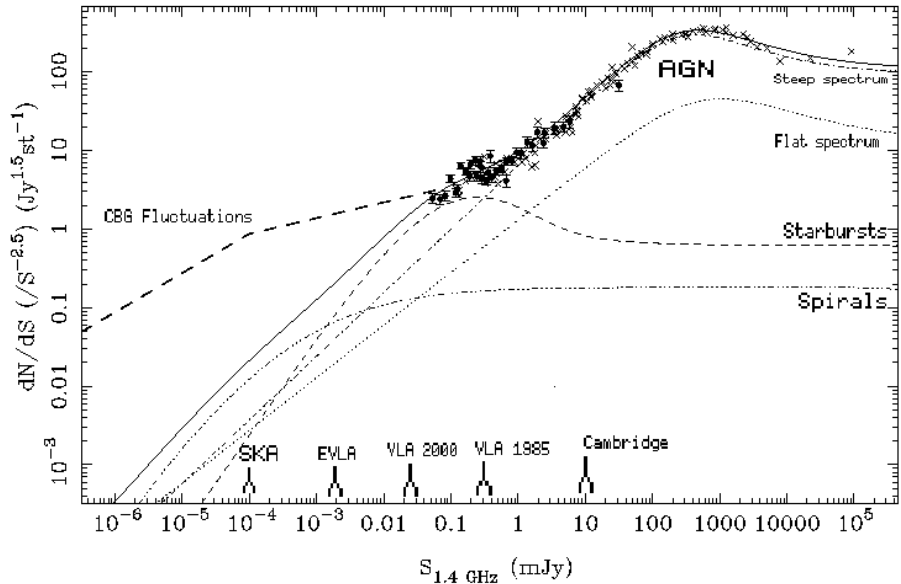,width=13cm}
\vskip -0.3in
\caption{Radio source counts at 1.4 GHz (adopted from 
\cite{hopkins2}).
}
\end{figure}    

Since deep radio images are usually  follow-up observations of
deep fields at other wavebands, there has been very rapid 
progress in determining the nature of the parent galaxies of
the $\mu$Jy radio source population \cite{richards,haarsma}.
This short summary will focus on recent results on the 
$\mu$Jy radio source population, emphasizing the unique information
about high redshift star forming galaxies 
coming from radio observations. I will also discuss the radio-to-far
infrared (FIR) correlation, re-deriving star formation rates
based on this correlation, and conclude with a short
discussion of future instrumentation.

Richards \cite{richards} finds that 75$\%$ of the 
S$_{1.4} \ge 40\mu$Jy radio sources are identified to
$I < 24$, with a median of $I = 22.1$. Interestingly,
he also finds that 25$\%$ of the sources are unidentified
to $I > 25$. And perhaps more interestingly, a number of groups
\cite{richards,barger99,bertoldi1,chapman} find that (sub)mm
observations of these $\mu$Jy radio sources with  faint 
(or absent) optical counterparts results in a $> 40\%$ detection
rate at mJy levels. This suggests that the optically faint, 
$\mu$Jy radio sources are equivalent to the mJy
(sub)mm source population. 

Haarsma et al. \cite{haarsma} have used an extensive spectroscopic and
photometric  
redshift analysis to determine the redshift distribution of
the $\mu$Jy radio source population. They find that (roughly):
\begin{itemize}
\item 50$\%$ of the sources are spirals, or irregular galaxies, at $z < 1$.

\item 25$\%$ are ellipticals,
presumably low luminosity AGN, at $0.3 < z < 1.5$.

\item 25$\%$ are optically faint (or absent), and red.
\end{itemize}

\noindent They propose that these later sources are likely 
to be high redshift ($1.5 < z < 4$), dust obscured  starbursts.
This idea is consistent with the SCUBA and MAMBO results
discussed above \cite{richards,barger99,bertoldi1,chapman}. 

The angular size distribution of the $\mu$Jy radio source population
remains a point of debate. WSRT observations of the HDF at 1.4 GHz,
15$''$ resolution, to 8$\mu$Jy rms detect a number of
sources not detected in the VLA survey at similar sensitivity but with
1$''$ resolution \cite{garrett}.  This would suggest a
significant population of sources larger than 1$''$. However,
confusion is a serious issue at this sensitivity level at
15$''$ resolution. Combined MERLIN+VLA observations of the HDF suggest
that most 
of the sources have angular sizes between 0.7$''$ and 2$''$, with a
median of 1.4$''$ (Muxlow et al. in prep). On the other hand, VLA
observations of the cluster A2125 suggest that most of the sources are
unresolved, with upper limits of typically 1$''$\cite{owen}.  If the
sources are 
indeed 1$''$ in size, this presents a significant problem for the SKA,
since the sky will become naturally confusion limited at the few nJy
level, independent of the resolution of the instrument
\cite{hopkins}, well above the sensitivity of the SKA.

Richards \cite{richards}
finds that the mean spectral index for a 1.4 GHz selected
sample of sources  
is --0.8, typical of star forming galaxies. Not surprisingly,
an 8 GHz selected sample shows a flatter mean spectral index of
--0.4. 

Barger et al. \cite{barger99} find that only 15$\%$ 
of the radio sources with S$_{1.4} \ge 25 \mu$Jy
in the Hawaii Deep Field are
X-ray sources with 2 to 10 keV fluxes of: 
$\rm I_x \ge 1\times10^{-15}$ erg s$^{-1}$ cm$^{-2}$.
This result is  consistent with the idea that the majority of
$\mu$Jy radio sources are star forming galaxies.


One problem with deep radio surveys is the limited area covered, such
that cosmic variance can lead to substantial differences between
counts derived for different fields. For instance, the
counts at the 100 $\mu$Jy level in the Phoenix Deep Field  
\cite{hopkins} are a factor two higher than those in the Hubble Deep
Field \cite{richards}. This variance can be seen as the increased scatter 
below 1 mJy in Fig. 1.  

\section{The FIR-radio correlation: deriving star formation rates
from radio observations and the importance of 
inverse Compton losses}

Most recent derivations of star formation rates
based on radio observations \cite{haarsma,cram,mobasher}
use equs. 21 and 23 in Condon \cite{condon}
to relate the star formation rate (SFR) to the 1.4 GHz spectral
luminosity. This relationship was derived
from the supernova rate and the integrated radio
luminosity of the Milky Way. 

An alternate method for deriving the relationship between
radio luminosity and SFR relies on the tight 
correlation between FIR luminosity and radio luminosity in 
star forming galaxies \cite{condon}. This correlation is remarkable in
the small 
scatter observed over at least three orders of magnitude
in FIR luminosity. Further, it holds for optical and IR selected 
samples \cite{yin}. This alternate method uses 
spectral synthesis models for star forming galaxies
\cite{leitherer}, assuming a fraction, $f_c$, of the bolometric
luminosity 
is absorbed by dust and re-emitted  in the infrared, and 
then uses the  FIR-radio correlation to relate the radio
luminosity to the FIR luminosity. 
For a 10$^8$ yr continuous starburst, solar abundances, 
a Salpeter IMF from 0.1 to 100 M$_\odot$, 
and $f_c$ = 1, this calculation leads to:

$$ \rm SFR = 5.1\times10^{-22}~~  L_{1.4}~~ M_\odot ~yr^{-1} ~~~~~~~~~(1)$$

\noindent where L$_{1.4}$ is the 1.4 GHz spectral luminosity in W
Hz$^{-1}$.  Scaling to the same  IMF limits, this equation 
implies a factor 2.5 lower SFR than the equations in  Condon
\cite{condon}.  

There are a number of uncertainties in both
calculations. In the case of Condon's calculation the Galactic
supernova rate and radio continuum luminosity are both uncertain by
at least 50$\%$.  For the stellar
synthesis model calculation, the scatter in the FIR-radio correlation
leads to a 50$\%$ uncertainty, while changing parameters in the
starburst model changes the predicted SFR, eg.
decreasing the starburst age to 10$^7$ years increases the SFR by
50$\%$, and there is the uncertain $f_c$. 
It is currently not clear which, if either, calculation is
correct. Indeed, the relative agreement is encouraging, given the
very different metods. And given the different 
conditions in different galaxies (eg. the age of the starburst,
the dust covering factor, the IMF, ...), it is clear that there
will be no globally correct relationship, only a statistically most
likely one.


The FIR-radio correlation has a very simple heuristic explanation:
both the FIR and radio emission relate to massive star
formation, with the FIR coming from dust heated by 
the interstellar radiation field, 
and the relativistic electrons being accelerated in
supernova remnant shocks. However, given the number of 
processes and parameters involved, it remains remarkable, and as yet 
unexplained, as to why the correlation is so tight
\cite{condon,duric}. For instance, it has long been known that
the total radio luminosity of 
galaxies, both normal disks and nuclear starbursts, is an order
of magnitude larger than that expected from the sum of the
supernovae \cite{condon,pooley}, although see \cite{colina}.
This requires that the relativistic electrons be stored in the ISM of
galaxies for a timescale, $\rm t \ge {{U_{B,SNR}}\over{U_{B,ISM}}}
\times t_{SNR} \sim 10^7 yr$, where U$_{\rm B}$ is the magnetic energy 
density. 

This leads to the question of 
the importance of inverse Compton losses off the microwave
background for relativistic electrons in the ISM of high redshift 
galaxies. The energy
density in the microwave background increases as:
$\rm U_{MWBG} = 4.0\times10^{-13} (1 + z)^4~ erg~ cm^{-3}$.
This is shown in Fig. 2,  along with the energy density
in the magnetic field in a typical spiral arm \cite{beck}.
The energy density in the magnetic field is 
larger than that in the microwave
background to $z \sim 1$. Beyond this redshift, inverse Compton
cooling will dominate over synchrotron radiation, limiting
electron lifetimes, t$_{1.4}$, and leading to a departure from the
radio-FIR relation. 

\begin{figure}
\hskip 0.6in
\psfig{figure=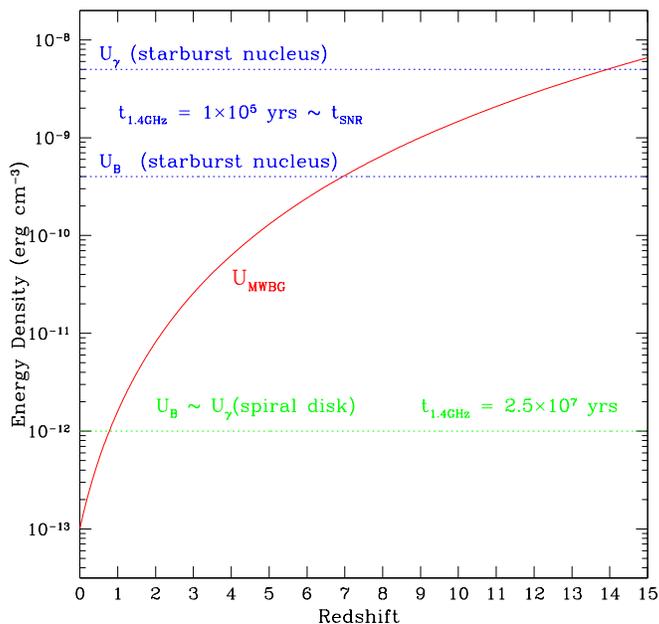,width=9cm}
\caption{The energy density of the microwave background, U$_{\rm
MWBG}$,  vs. redshift.
Also shown is the typical magnetic energy density, $\rm U_B$, in
the arm of a spiral galaxy,  and $\rm U_B$ in compact nuclear
starbursts, and 
the energy density in the radiation field, $\rm U_\gamma$,
in starbursts. t$_{1.4}$ is the radiative lifetime of an
electron emitting at 1.4 GHz.
}
\end{figure}

This is not true, however, for compact
nuclear starburst galaxies, ie. systems with SFRs
$> 100$ M$_\odot$ year$^{-1}$ in regions smaller than a few hundred
parsecs \cite{Solomon}. In these systems the  energy
density in the magnetic field is thought to be almost three orders of
magnitude larger than in the disk \cite{Taylor99}, in which case 
$\rm U_{MWBG}$ only becomes relevant at $z > 7$.
But compact nuclear starbursts raise a different, related problem:
the energy density in the
IR radiation field from the starburst itself is larger still than that
in the  magnetic field. This means that inverse Compton cooling should 
remove the synchrotron emitting electrons on fairly short timescales
($\le 10^5$ yr), and accentuates the question: why do
nuclear starbursts follow the radio-FIR correlation?  
There is a large body of literature on this issue \cite{condon},
but as yet no closure.  Hence, we trade one problem, inverse Compton
losses off the microwave background at high $z$, for a second,
inverse Compton losses off the starburst IR radiation field at all
$z$. 

Until proven otherwise, we adopt the radio-FIR correlation as a 
given, and see how it can be used in the study of high $z$ star
forming galaxies. 

\section{Using the radio-FIR correlation to study high $z$ star
forming galaxies, and what the future holds}

Haarsma et al. \cite{haarsma} have derived the cosmic 
star formation rate density (SFRD) vs. $z$  based 
on  $\mu$Jy radio samples. They find a steep rise in the 
density from $z = 0$ to  $z = 1$,  as has been found in optical and
IR studies. But they also find a systematically higher 
SFRD at all redshifts by a factor 3 relative to reddening corrected
optical studies.  This suggests that even
larger dust corrections are needed in optical studies, or that
optical studies miss a large population of dust-obscured galaxies. 
However, they use the equations in Condon \cite{condon} to
derive SFRs from radio continuum luminosities.
If we use equ. 1 above instead,  the radio derived values agree
well with the optical values.

A second area in which the radio-FIR correlation has been used in the
study of high $z$ star forming galaxies is as a redshift indicator
\cite{yun99,yun00,blain,dunne,low,hughes,barger99}.
The impetus in this case is the very faint optical counterparts being
found for most faint (sub)mm sources, thereby precluding
follow-up optical spectroscopy of a large sample of
sources. The radio-FIR method relies on the opposing slopes of the
synchrotron and thermal dust emission in star forming galaxies.  Our
most recent models use the average observed SED for nearby starbursts to 
relate redshift to the observed spectral index between 350 GHz and 1.4
GHz \cite{yun00}. Figure 3 shows 
the model, along with a few sources with known redshifts, including 
sources with AGN and starburst optical spectra.  The method is
admittedly imprecise, especially at high redshift, and 
there are a few degeneracies in the solutions, 
such as the addition of cold dust or the presence of a radio loud AGN 
\cite{blain}, but it is the
only viable alternative for deriving redshifts for the large majority
(90$\%$) of the faint (sub)mm sources.
The redshift distribution for the faint (sub)mm source population as
derived using the curves from \cite{yun00} can found in the
contribution by Bertoldi in this volume. 

\begin{figure}
\hskip 0.6in
\psfig{figure=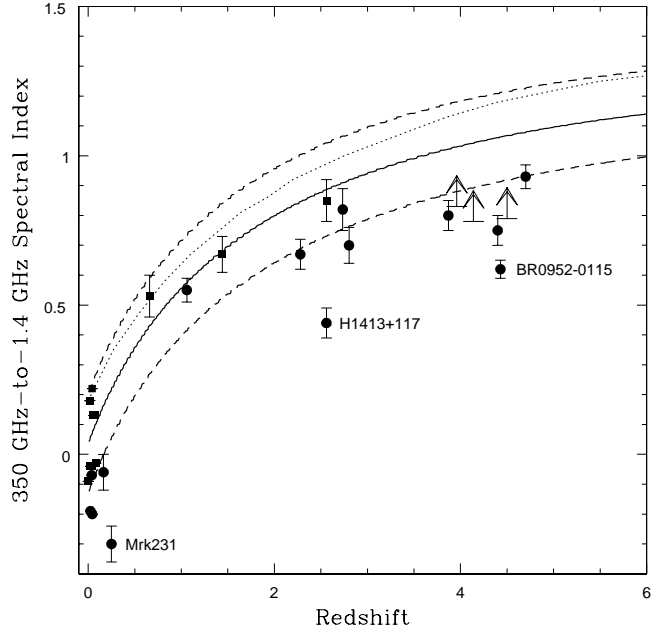,width=9cm}
\caption{The relationship
between redshift and observed 1.4-to-350 GHz
spectral index  for an active star forming galaxy, 
derived from the observed SEDs of 17 low $z$ galaxies
\cite{yun00}. The dash line shows the \cite{dunne}
model. Squares are observed values for galaxies with starburst
spectra. Circles are values for sources with AGN spectra.
}
\end{figure}

An important point concerning radio follow-up observations of
faint (sub)mm sources is the relative sensitivities. 
Comparing MAMBO images with  deep radio images shows that  
70$\%$ (10 of 14)  of the $\ge 3.5$ mJy (5$\sigma$)
sources at 250 GHz have radio counterparts with $\rm S_{1.4} \ge 
22 \mu$Jy within 3$''$ (Bertoldi, this volume).   We expect
only 0.2  chance  coincidences.  
This result lends confidence to the reality of the
MAMBO detections. It also implies that the current sensitivity of deep
VLA fields is well matched to that obtained with
mm bolometer arrays, and in particular, that there is not
a dominant population of very high $z$ sources ($z > 4$), or of low
$z$, cold dust sources.

Figure 4 shows the expected sensitivity of future cm to IR instruments
compared to the expected flux density of Arp 220 at various redshifts.
Overall, the next generation instruments are well matched to
the expected flux density of Arp 220 out to $z \sim 2~\rm to~ 8$.
Clearly, the ALMA is by far the most sensitive telescope 
relative to the
dust spectrum, and will detect low luminosity galaxies to
very high $z$. However, this applies to pointed observations, which
have a limited field-of-view (FWHM $\sim 20''$).  For surveys of fields
larger than about $15' \times 15'$, the next generation bolometer array 
cameras operating on large single dish telescopes 
will be competitive with ALMA, as will the EVLA and SIRTF. 

Figure 5 shows the
flux density of Arp 220 and M82 vs. $z$ 
at various frequencies, relative to the sensitivities of
future telescopes. 
The important point in this diagram is the interesting
source selection function at (sub)mm wavelengths: 
typical (sub)mm surveys
detect luminous star forming galaxies at essentially
all redshifts, but they miss completely the low $z$,
low luminosity galaxies. Hence (sub)mm surveys result in a very clean, 
but totally biased, sample of sources. 
Radio observations result in a mixture
of low $z$, low luminosity, and high $z$, high
luminosity star forming galaxies, as well as radio loud AGN.

The closing debate at this workshop contrasted the IR vs. submm
vs. optical views of high $z$ galaxies and galaxy formation.
Each side argued that they detect the dominant contribution to
cosmic star formation at a specific epoch. 
Yet, each side has a very specific 
galaxy selection function, and the overlap between the populations
apparently is small, $\sim 10\%$. 
Indeed, one might argue that $\mu$Jy radio
surveys are the least biased means of detecting all the 
source populations. But therein lies the fundamental problem:
how to differentiate the source populations on a deep radio image?
Overall, it is clear that all sides are currently seeing only
limited, and perhaps orthogonal, aspects of galaxy formation. In order
to address the general question of  
galaxy formation, or at least the formation of the stellar content of
galaxies, requires  wide field surveys
using the EVLA, the next generation bolometers arrays on the LMT and
GBT, and SIRTF, with very deep follow-up studies 
of selected samples of sources using ALMA and NGST. 

The National Radio Astronomy Observatory is
operated by Associated Univ. Inc., under contract with the
National Science Foundation. I would like to thank A. Hopkins
for allowing me to reproduce Fig. 1, and B. Poggianti, N. Miller,
A. Blain, M.Yun, and F. Owen for useful discussions. 

\begin{figure}
\hskip 0.6in
\psfig{figure=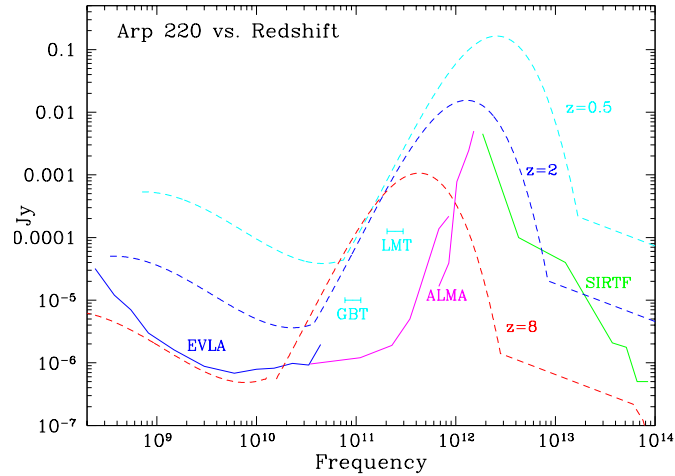,angle=-90,width=9cm}
\caption{The observed spectrum of Arp 220 
at various redshifts, compared to the  sensitivities for various
existing and future telescopes. 
}
\end{figure}    

\begin{figure}
\vskip -0.5in
\hspace*{-0.4in}
\psfig{figure=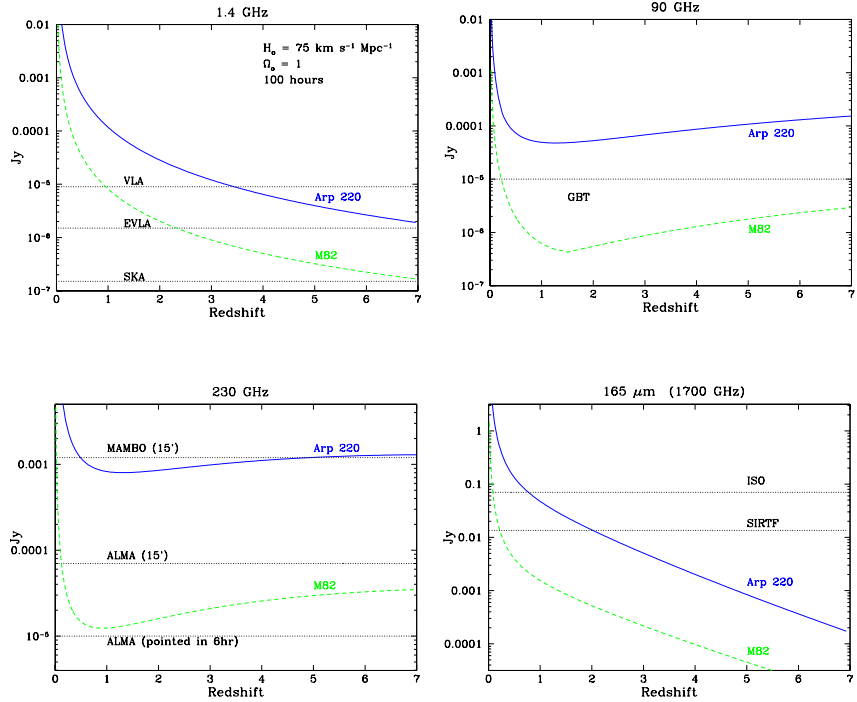,angle=-90,width=14cm}
\caption{The flux densities of Arp 220 and M82 vs. $z$
at various observing frequencies.
Also shown are the sensitivities of existing and
future telescopes. 
}
\end{figure}

\end{document}